\documentclass[english, 12pt]{article}
\def\DEFAULTFONT{}

\usepackage{parskip}
\usepackage{base}
\usepackage{macros}
\usepackage{minimacros}
\DeclareMathOperator{\ord}{ord}
\DeclareMathOperator{\LCM}{LCM}
\algtext*{EndIf}
\algtext*{EndFor}
\algtext*{EndProcedure}

\title{Deterministically finding an element of large order in $\mathbb{Z}_N^*$}
\author{Itamar Nir\thanks{This research was funded by the European Union (ERC, EACTP, 101142020). Views and opinions expressed are however those of the author(s) only and do not necessarily reflect those of the European Union or the European Research Council Executive Agency. Neither the European Union nor the granting authority can be held responsible for them.}}
\date{ }

\begin{document}

\maketitle

\begin{abstract}
In this paper, we present an improvement for the problem of deterministically finding an element of large multiplicative order modulo some integer $N$. 
This problem arises as a key subroutine in current deterministic factoring algorithms, such as those proposed by Harvey
and Hittmeir [Mathematics of Computation, 2021].

Specifically, let $D<N$ be positive integers with 
\begin{equation}\label{eq:abs} D > \exp\left(\sqrt{2\log N \log \log N}\right). 
\end{equation} 
We give a deterministic algorithm that does one of the following: Returns an element $a \in \mathbb{Z}_N^*$ with $\operatorname{ord}_N(a) > D$; Returns a non-trivial factor of $N$;
Or reports that $N$ is prime.
The running time of our algorithm is $O(D^{1/2 + o(1)})$. Similar results were independently and concurrently obtained by Harvey and Hittmeir [arXiv:2601.11131, 2026] in work that appeared while this manuscript was in preparation. Prior to these works, the best known algorithm for finding an element with order larger than $D$ was given by Oznovich and Volk [SODA 2026], requiring $D > N^{\frac{1}{6}}$. We also present a simpler algorithm that applies for any $D < N$ and runs in $O(D^{2.5+o(1)}\operatorname{polylog}(N))$.
\end{abstract}

\tableofcontents

\section{Introduction}

For a positive integer $N$, and a positive integer $a$ coprime to $N$, we denote by $\ord_N(a)$ the smallest positive integer $k$ for which $a^k \equiv 1 \pmod N$. Given an upper bound $D$, a theorem of Sutherland~\cite{sutherland2007order}
computes $\ord_N(a)$ in $O(D^{1/2+o(1)})$ time when $\ord_N(a) \le D$,
and otherwise reports $\ord_N(a) > D$. The problem we address in this paper is the following: given $N$ and a positive integer $D$, deterministically find an integer $a$ satisfying $\ord_N(a)> D$ while maintaining the same time complexity of $O(D^{1/2+o(1)})$. 

Finding an element of large order is closely related to integer factorization. For about 45 years, the fastest known deterministic factoring algorithm was due to Strassen~\cite{strassen1977einige}, which ran in $\tilde{O}(N^{1/4})$. A recent breakthrough by Hittmeir~\cite{hittmeir2021timespace} improved the complexity to $\tilde{O}(N^{2/9})$, which was subsequently improved to $\tilde{O}(N^{1/5})$ by Harvey~\cite{harvey2021exponent}. Both of these recent algorithms rely on finding an element of large order, highlighting the relevance and importance of this problem.

A small challenge that arises when addressing this task is that an element of large enough order might not exist. However, if such an element does not exist, we will demonstrate that $N$ must have a small prime factor, which we will be able to find efficiently. Thus, our goal is to either find an element of order larger than $D$, or find a factor of $N$ if $N$ is composite. If $N$ is prime, we simply report this fact. 

If $N$ has no small prime factors, the problem is easy with randomness:
a uniformly random $a \in \Z_N^*$ has $\ord_N(a) > D$ with high
probability, and \cref{thm:order-calculation-alg} verifies this. However, this problem turns out to be much harder when a rigorously proven deterministic algorithm is required.

Prior to this work, the best known algorithm for finding an element of order larger than $D$ is due to Oznovich and Volk~\cite{oznovich2026deterministic}. Their method achieves the running time of $O(D^{1/2 + o(1)})$ and applies for $D > N^{1/6}$.

\subsection{Our results}
In this paper, we relax this condition of $D > N^{1/6}$ to $D$ satisfying \cref{eq:abs}. Specifically, we prove the following theorem:
\begin{theorem}\label{thm:main}
There exists an algorithm that receives as input positive integers $D<N$ with $D > \exp\left(\sqrt{2\log N \log \log N}\right)$, and returns one of the following in $O(D^{1/2 + o(1)})$ time:
\begin{enumerate}
    \item An element $a \in \mathbb{Z}_N^*$ with $\ord_N(a) > D$;
    \item A non-trivial factor of $N$;
    \item "$N$ is prime".
\end{enumerate}
\end{theorem}

We also give a simpler deterministic algorithm with the same output guarantee
for arbitrary \(D<N\), but with a weaker running time:

\begin{proposition}
\label{prop:slow-algorithm}
There exists a deterministic algorithm that, given positive integers $D<N$,
returns in time $O(D^{2.5+o(1)} \operatorname{polylog}(N))$ one of the
following:
\begin{enumerate}
    \item An element $a \in \mathbb{Z}_N^*$ with $\ord_N(a) > D$;
    \item A non-trivial factor of $N$;
    \item ``$N$ is prime''.
\end{enumerate}
\end{proposition}
This algorithm may be useful for small values of \(D\). It is simpler than the
main algorithm, while still sharing many of its core ideas.

\subsection{Main algorithm overview}
Our algorithm follows the general framework introduced by Hittmeir~\cite{hittmeir2018babystep}. One first scans
small integers \(a\). If such an \(a\) has order larger than \(D\), we are done. Otherwise,
we compute \(\ord_N(a)\). Using \cref{lem:order-all-primes}, we either obtain a nontrivial
factor of \(N\), or reduce to the case where the computed order of \(a\) modulo \(N\) is
also its order modulo every prime divisor of \(N\).

Let \(M\) be the least common multiple of the orders computed in this way. In the latter
case, for every prime divisor \(p\) of \(N\), each of these orders divides \(p-1\), and hence
\[
M \mid p-1.
\]
Thus the remaining task is to exploit the congruence \(p\equiv 1 \pmod M\) in order to find
a prime divisor \(p\) of \(N\), or to prove that \(N\) is prime.

Previous algorithms, such as those of Hittmeir~\cite{hittmeir2018babystep} and Oznovich
and Volk~\cite{oznovich2026deterministic}, use more involved procedures at this stage.
The main observation of this paper is that, under the assumption
\[
D>\exp(\sqrt{2\log N\log\log N}),
\]
a naive search over the integers congruent to \(1\pmod M\) is already fast enough. The
reason is that the computed orders force \(M\) to be large: many \(\sqrt D\)-smooth integers
become roots of \(x^M-1\) modulo the smallest prime divisor of \(N\).

This yields a particularly simple \(O(D^{1/2+o(1)})\)-time algorithm in the above range of
\(D\).

\subsection{Independent work}
In concurrent and independent work, Harvey and Hittmeir~\cite{harvey2026deterministic}
use the same general order-computation framework, but proceed differently after the
orders of small integers have been computed. Their method often constructs an element
of large order directly. This leads to an \(O(D^{1/2+o(1)})\)-time algorithm for all
\(D<N\). While our contribution is slightly more limited in range, the resulting algorithm and analysis are notably simpler.

\section{Preliminaries}

In this section, we cite several important results that we will use in our algorithm and analysis.

The first result is an algorithm by Sutherland, which allows us to find the order of an element given an upper bound on its order.

\begin{theorem}[\cite{sutherland2007order}]
    \label{thm:order-calculation-alg}
    There exists an algorithm that, given inputs $D, N \in \N$ such that $D<N$ and $a \in \Z_N^*$, runs in time
    \[
    O\left(\frac{D^{1/2}}{\sqrt{\log \log D}} \cdot \log^2 N \right),
    \]
    and outputs:
    \begin{itemize}
        \item $\ord_N(a)$ if $\ord_N(a) \leq D$
        \item ``\( \ord_N(a) > D \)'' otherwise.
    \end{itemize}
\end{theorem}

The next result by Pomerance provides a lower bound on the density of smooth numbers.

\begin{theorem}[\cite{KP97}]\label{smooth numbers}
    Let $\psi(x,y)$ denote the number of $y$-smooth integers up to $x$, i.e., integers whose prime factors do not exceed $y$. If $2 \le y \le x$ and $x \ge 4$, then
    \[
    \psi(x,y) \ge x^{1 - \frac{\log \log x}{\log y}}.
    \]
\end{theorem}

Finally, we use the following lemma to argue that if an element $a \in \mathbb{Z}_N^*$ has order $m$, then either $m$ is also its order modulo any prime divisor of $N$, or we can factor $N$.  

\begin{lemma}[\cite{hittmeir2018babystep}]
    \label{lem:order-all-primes}
    Let \( N \in \mathbb{N} \) and $a \in \Z_N^*$. Let \( m := \ord_N (a) \).
    Then
    \[
        \ord_p(a) = m \quad \text{for every prime } p \text{ dividing } N
    \]
    if and only if
    \[
        \gcd(N, a^{m/r} - 1) = 1 \quad \text{for every prime } r \text{ dividing } m.
    \]
\end{lemma}

\section{A simple algorithm for arbitrary $D$}

In this section, we prove \cref{prop:slow-algorithm}.

We go through the integers
\[
2,3,\ldots,D^2+D.
\]
For each such integer \(a\), we first check whether \(a\mid N\). If so, we
return \(a\) as a nontrivial factor of \(N\) (or report that \(N\) is prime if
\(a=N\)). Otherwise, \(a\) is coprime to \(N\): indeed, if \(\gcd(a,N)>1\),
then some prime \(q\mid a\) also divides \(N\), and since \(q\le a\), we would have already returned when we reached \(q\). Thus we may apply
\cref{thm:order-calculation-alg} to compute \(\ord_N(a)\), or else conclude
that \(\ord_N(a)>D\).

We claim that if this algorithm does not find a divisor of \(N\) among the
integers \(2,3,\ldots,D^2+D\), then one of these integers must have order
larger than \(D\). Indeed, suppose otherwise. Then every
\(a=2,3,\ldots,D^2+D\) satisfies \(\ord_N(a)\le D\). Let \(p\) be a prime
divisor of \(N\). Since no divisor of \(N\) was found in this range, we have
\(p>D^2+D\), and hence the integers \(2,3,\ldots,D^2+D\) are distinct modulo
\(p\). Moreover,
\[
\ord_p(a)\le \ord_N(a)\le D
\]
for each such \(a\). By the pigeonhole principle, there exists
\(1\le k\le D\) such that at least \(D+1\) of these integers satisfy
\[
a^k\equiv 1 \pmod p.
\]
This contradicts the fact that the nonzero polynomial \(x^k-1\) has at most
\(k\le D\) roots in \(\mathbb F_p\). Therefore, unless the algorithm finds a
factor of \(N\), it must find an element \(a\in\mathbb Z_N^*\) with
\(\ord_N(a)>D\).

The running time is \(O(D^{2.5+o(1)}\operatorname{polylog}(N))\), since the
algorithm performs \(O(D^2)\) iterations, and each call to
\cref{thm:order-calculation-alg} costs \(O(D^{1/2+o(1)}\operatorname{polylog}(N))\).

This algorithm has many ideas in common with our main algorithm: it scans
small integers, either finds a divisor of \(N\), or uses order computations to
certify that one of them has order larger than \(D\). The main algorithm can be
viewed as a faster version of this idea, using the least common multiple of the
computed orders and a smooth-number estimate to reduce the search range.

\section{Main algorithm}

In this section, we prove \cref{thm:main}. We first present the claimed algorithm, then prove its correctness, and finally show that it runs in $O(D^{1/2 + o(1)})$ time.

\begin{algorithm}[H]
    \caption{Finding an Element with Large Order or a Nontrivial Factor of \(N\)}
    \label{alg:algorithm}
    \raggedright
    Input: Integers $N, D$ satisfying $D > \exp\left(\sqrt{2\log N \log \log N}\right)$ and $D < N$\\
    Output: An element \(a \in \mathbb{Z}_N^*\) with \(\ord_N(a) > D\), or a nontrivial factor of \(N\), or ``$N$ is prime''.
    \noindent\rule{\textwidth}{0.4pt}
    \begin{algorithmic}[1]
        \If{$N < 100$}
            \State Return the correct answer using brute force.
        \EndIf
        \State Set \(M = 1\) \label{step:first-line}
        \For{$a = 2, 3, \dots, \lceil\sqrt{D}\rceil$} \label{step:main-loop}
            \If{\(a^{M} \equiv 1 \pmod{N}\)} \label{step:while-a-order-div-M}
                \State Go to the next $a$
            \EndIf
            \If{\(a \mid N\)}
                \State Return \(a\) as a nontrivial factor of \(N\) or, if \(a = N\), ``$N$ is prime''.\label{step:a-div-N-return}
            \EndIf 
            \State Run the algorithm of~\cref{thm:order-calculation-alg} on \( D, N\) and $a$  \label{step:BSGS}
            \If {"$\ord_N(a)>D$" was returned}
                \State Return \(a\) as element with \(\ord_N(a) > D\) \label{step:large-order-a-found}
            \EndIf
            \State Set \(m_a = \ord_N(a)\) and compute the prime factorization of $m_a$ using the naive $\tilde{O}\br{\sqrt{m_a}}$ algorithm. \label{step:calculate-order}
            \For{each prime \(q\) dividing \(m_a\)} \label{step:foreach-prime-dividing-me}
                \If{\(\gcd(N, a^{m_a/q} - 1) \neq 1\)} \label{step:compute-gcd}
                    \State Return \(\gcd(N, a^{m_a/q} - 1)\) as a nontrivial factor of \(N\) \label{step:gcd-divides-N}
                \EndIf
            \EndFor
            \State Set \(M \gets \LCM(M, m_a)\) \label{step:update-M}
        \EndFor
        \For{$k = 1, 2, \dots$}\label{second loop}
            \If{$kM+1 > \sqrt{N}$}
                \State Return ``$N$ is prime''
            \EndIf
            \If{$kM+1 \mid N$}
                \State Return $kM+1$ as a non-trivial factor of $N$ \label{step:p=1mod M-found}
            \EndIf
        \EndFor
    \end{algorithmic}\label{alg:main}
\end{algorithm}

\subsection{Correctness}
We first note that when line \ref{step:BSGS} is reached, $a$ and $N$ are indeed coprime. Any integer $g >1$ such that $g \mid a,N$ would have been returned on the current or on an earlier iteration of line \ref{step:a-div-N-return}.

It is clear that if the algorithm terminates on lines \ref{step:a-div-N-return} or \ref{step:large-order-a-found} then the answer is correct. 
Now suppose the algorithm returned on line \ref{step:gcd-divides-N}. Since $\ord_N(a) = m_a$, $N$ must not divide $a^{\frac{m_a}{q}} - 1$. Therefore, $\gcd(N, a^{\frac{m_a}{q}} - 1)$ is indeed a non-trivial divisor of $N$.

Now assume that line \ref{second loop} is reached. Since we never returned on line \ref{step:gcd-divides-N}, Lemma \ref{lem:order-all-primes} implies that $m_a$ is the order of $a$ modulo each of the prime factors of $N$. In particular, $m_a \mid p-1$ for every prime $p \mid N$. Since $M$ is the LCM of those $m_a$'s, $M \mid p-1$ also holds for all $p \mid N$.
Thus, either line \ref{step:p=1mod M-found} will be reached and we will return the smallest prime factor of $N$, or "$N$ is prime" will be returned if this is the case.

\subsection{Running time analysis}
First, observe that \cref{eq:abs} implies \(\log N = D^{o(1)}\). Hence any
factor that is polylogarithmic in \(N\) can be absorbed into the \(D^{o(1)}\)
term in our running-time bounds.

Lines \ref{step:while-a-order-div-M}-\ref{step:a-div-N-return} are executed at most $\lceil\sqrt{D}\rceil$ times, and take polylogarithmic time in $N, D$.
Note that in the iterations where line \ref{step:BSGS} is reached, $m_a$ must not divide $M$. Therefore, $\LCM(M, m_a) \ge 2M$, so $M$ is multiplied by at least 2 in line \ref{step:update-M}. Since every $m_a$ divides $\varphi(N)$, so does their least common
multiple $M$; hence $M \le \varphi(N) < N$, so this can happen at
most $\log_2 N$ times. So lines \ref{step:BSGS}-\ref{step:update-M} will be executed at most $\log N$ times.

Line~\ref{step:BSGS}  runs in $O(D^{1/2 + o(1)})$ by Theorem \ref{thm:order-calculation-alg}. Line \ref{step:calculate-order} also takes $O(D^{1/2+o(1)})$ time since $m_a \le D$. All other lines take polylogarithmic time in $N, D$.
To sum up, since lines \ref{step:first-line}-\ref{step:a-div-N-return} take polylogarithmic time in $N$ and are executed $O(\sqrt{D})$ times, and lines \ref{step:BSGS}-\ref{step:update-M} take $O(D^{1/2 + o(1)})$ time and are executed at most $\log N$ times, the running time of lines \ref{step:first-line}-\ref{step:update-M} is $O(D^{1/2 + o(1)})$.

All that remains to do is bound the number of iterations that the loop in line \ref{second loop} performs.

Let $p$ be the smallest prime divisor of $N$. According to our correctness analysis, if the algorithm doesn't return before line \ref{second loop}, $p\equiv 1 \pmod M$ is satisfied. Thus, if $N$ is composite, $p$ is found in exactly $\frac{p-1}{M}$ iterations; if $N$ is prime, we return "$N$ is prime" after $\lceil\frac{\sqrt{N}}{M}\rceil$ iterations. Let $L := \min(\lceil{\sqrt{N}\rceil},p)$. Since $p \le\sqrt{N}$ when $N$ is composite, at most $\lceil\frac{L}{M}\rceil$ iterations are performed in both cases. We will prove that $\frac{L}{M} \le O(\sqrt{D})$. 

Observe that by definition, $a^M \equiv 1 \pmod{N}$ holds for all $a = 1, 2, \dots, \lceil\sqrt{D}\rceil$. In particular, $1, 2, \dots, \lceil\sqrt{D}\rceil$ are all roots of the polynomial $x^M - 1$ over \(\mathbb F_p\). Note that $p>\lceil\sqrt{D}\rceil$ since $1,2,...,\lceil\sqrt{D}\rceil$ are coprime to $N$. As a consequence, $1, 2, \dots, \lceil\sqrt{D}\rceil$ are all distinct modulo $p$. We already obtain the bound $M \ge \sqrt{D}$, since a nonzero polynomial over a field has at most as many roots as its degree.

Furthermore, note that if $a^M \equiv 1 \pmod{p}$ and $b^M \equiv 1 \pmod{p}$, then $(ab)^M \equiv 1 \pmod{p}$. This leads us to the observation that all $\sqrt{D}$-smooth numbers are also roots of $x^M - 1$ over \(\mathbb F_p\). Hence, there are at most $M$ $\sqrt{D}$-smooth numbers up to $p$.

We verify that the hypotheses of \cref{smooth numbers} hold for the parameters \(x=L\) and \(y=\lceil\sqrt D\rceil\) given that the
second loop is reached. We know that $\sqrt{N} > \sqrt{D}$ and $p > \sqrt{D}$ so $L = \min(\lceil\sqrt{N}\rceil,p) \ge \lceil\sqrt{D}\rceil$. In addition, we know $N \ge 100$, so by \cref{eq:abs}, we have $D \ge 10$, and $p \ge \sqrt{D} > 3$. So $L = \min(\lceil\sqrt{N}\rceil,p)\ge 4$.
Thus the parameters \(x=L\) and \(y=\lceil\sqrt D\rceil\) lie in the required range for
\cref{smooth numbers}.

By \cref{smooth numbers}
\[
M \ge \psi(p, \lceil\sqrt{D}\rceil) \ge \psi(L,\lceil\sqrt{D}\rceil) \ge L^{1 - \frac{\log \log L}{\log \sqrt{D}}}.
\]
Thus
\[
\frac{L}{M} \le L^{\frac{\log \log L}{\log \sqrt{D}}} = \exp\left(\frac{\log L\log \log L}{\log \sqrt{D}}\right)  \stackrel{(\star)}{\le} \ O(\sqrt{D}) = \exp(\log \sqrt{D}+O(1))
\]
where $(\star)$ uses the inequality
\[
\frac{\log L \log\log L}{\log\sqrt D} \;\le\; \log\sqrt D + O(1),
\tag{$\star$}
\]
which we now establish.
Indeed, using $L \le \lceil\sqrt{N}\rceil$ and \cref{eq:abs}: 
\[
\log L\log \log L\le \log \lceil\sqrt{N}\rceil \log \log \lceil\sqrt{N}\rceil \le\frac{1}{2} \log N \log \log N +O(1) \le \frac{1}{4} \log^{2} D+O(1) = \log^{2} \sqrt{D}+O(1)
\]
which proves $\frac{L}{M} \le O( \sqrt{D})$ as we wished.

Thus, the second loop performs \(O(\sqrt D)\) iterations, and the total running
time of the algorithm is \(O(D^{1/2+o(1)})\).

\section*{Acknowledgments}
The author wishes to thank their advisor, Prof. Amir Shpilka, for helpful discussions and guidance throughout this work. 

\bibliographystyle{alpha} 
\bibliography{references}

\end{document}